\documentclass[11 pt]{amsart}
\usepackage{amssymb,amsmath}
\newtheorem{theorem}{Theorem}[section]
\newtheorem{lemma}[theorem]{Lemma}
\newtheorem{corollary}[theorem]{Corollary}
\newtheorem{proposition}[theorem]{Proposition}

\newtheorem{definition}[theorem]{Definition}

\begin{document}

\title[A note on decidability of cellularity ]{A note on decidability of cellularity}
\author[U.B. Darji]{Udayan B.Darji}
\author[S.W. Seif]{Steve W. Seif}
\address{Department of Mathematics\\ University of Louisville\\
Louisville, KY 40292,USA}
\email{ubdarj01@louisville.edu}
\address{Department of Mathematics\\ University of Louisville\\
Louisville, KY 40292,USA}
\email{swseif01@louisville.edu}
\date{November 21, 2007}
\maketitle

\begin{abstract} A regular language $L$ is said to be cellular if there exists a $1$-dimensional cellular automaton
 $CA$ such that 
%% added the word ``the'' before ``finite blocks''
$L$ is the language consisting of the finite blocks associated with $CA$. It is shown that cellularity of a regular language is 
decidable using a new characterization of cellular languages formulated by Freiling, Goldstein and Moews and implied by
a deep result of Boyle in symbolic dynamics.
\end{abstract}

\section{Introduction}

A $1$-dimensional cellular automaton CA of \emph{radius} $r$ 
consists of a finite alphabet
 $\Gamma$ and a \emph{local rule} 
$\rho: \Gamma^{2r+1}\to \Gamma$.
 The local rule $\rho$ extends to a function $\rho_{\infty}: \Gamma^{\mathbb{Z}}\to \Gamma^{\mathbb{Z}}$ as follows: for 
$c \in \Gamma^{\mathbb{Z}}$ and $i\in \mathbb{Z}$, $\rho(c)_{\infty}(i) = \rho(c_{i-r}, \ldots, c_i, \ldots, c_{i+r})$. There is a vast literature about cellular automata; they have been used to model a wide variety of physical and mathematical phenomena. One aspect of the study of $1$-dimensional cellular automata involves comparing properties of $CA$ with those of a certain regular language, $L(CA)$, which is now described.
 A \emph{finite block} of $\rho_{\infty}(c)$ is a finite sequence of contiguous entries in $\rho_{\infty}(c)$, of the form $\rho_{\infty}(c) (m), \ldots ,\rho_{\infty}(c) (m+k)$, where $m \in \mathbb{Z}$ and $k\in \mathbf{N} \cup \{0\}$. As the range of $\rho_{\infty}$ is shift invariant, we may replace $m$ by $0$ in the definition of finite block. Let $L(CA)$ be $\{\rho_{\infty}(c) (0), \ldots ,\rho_{\infty}(c) (k): c\in \Gamma^{\mathbb{Z}}, \; k\in \mathbb{N} \cup \{0\}\} \cup \{\epsilon\}$
(where $\epsilon$ is the empty word). 
%%I added the next sentence and modified the references to Sutner. 
It is well known that $L(CA)$ is accepted by a non-deterministic automaton whose associated undirected graph is a De Bruijn graph. The reader may refer to Sutner's work \cite{Sutner} and \cite{Sutner2} for further information on $1$-dimensional cellular automata and regular languages. Kari \cite{kari} presents a broad
range of topics and problems connected with cellular automata.

\textbf{}

A regular language $L$ is said to be \emph{cellular} if there exists a $1$-dimensional cellular automaton $CA$ such that $L = L(CA)$. 
 Klaus Sutner has asked whether cellularity is decidable in the regular languages, and his question is answered in the affirmative here. 

\begin{theorem} \label{main result}
Cellularity is decidable.
\end{theorem}

%%Minor syntax changes in the parag. below

The algorithm constructed here depends on a very useful and apparently new characterization of cellular 
languages. The first author learned this characterization, given in Corollary~\ref{properties}, from Freiling, Goldstein and Moews who had proved
it in terms of cellular languages. 
%%I separated the last sentence and next one into 2 sentences.
 They later realized that it is implied by a deep result of Boyle \cite{Boyle} in symbolic
dynamics concerning mixing of sofic shifts. 
%%Darji, I don't think we need to say this, but I left it in case % you don't agree.
Notation
and terminology necessary for the statement of Corollary~\ref{properties} are given next.

All languages are assumed to be over a finite, non-empty alphabet $\Gamma$. As is customary we use $\Gamma ^*$ to denote
all the finite word over $\Gamma$, including the empty word $\epsilon$.
Because cellular languages are non-empty and properly contain 
$\{\epsilon\}$, all languages here will be assumed to have those 
properties. Since $\Gamma^*$ is cellular, it will also be assumed 
that $L$ is properly contained in $\Gamma^*$.

\vspace{.1 in}

\begin{definition} \label{properties} Let $L$ be a language over a 
finite alphabet $\Gamma$. Four properties are defined.
\begin{enumerate}
\item (L1)(Factorial) If $x,y \in \Gamma^*$ and $xy \in L$, then $x, y\in L$.
\item (L2)(Receptive) There exists a letter $o\in \Gamma$, such that for all 
$x,y\in L$, there exist $a,b\in L$ such that for all $n\in 
\mathbb{N}$, $xao^nby \in L$. 
\item (L3) (Transitive) If $x,y\in L$, there exists $u\in L$ such that $xuy\in 
L$.
\item (L4) (Prolongable) If $u\in L$, then there exist $s,t \in L - 
\{\epsilon\}$ such that $sut\in L$.
%%\item (L5) (Star-able) There exists $o\in \Gamma$ such that $o^* 
%%\subset L$.
\end{enumerate}
\end{definition}

It is not difficult to verify that (L1) through (L4) are satisfied by 
cellular languages. (L1), (L3), and (L4) are familiar to those who 
work in cellular automata and symbolic dynamics; (L2) is a condition 
that arises naturally from Boyle's Theorem 3.3 \cite{Boyle}. Note 
that (L2) implies both (L3) and (L4).

\vspace{.1 in}

\begin{corollary} \label{fgmb} (Freiling-Goldstein-Moews \cite{fgm}, Boyle \cite{Boyle}) A regular language $L$ over an alphabet 
$\Gamma$ is cellular if and only if $L$ satisfies (L1) and (L2). 
\end{corollary}

Some results and notation from basic automata theory are briefly 
reviewed. The reader may refer to \cite{Sipser} for further detail. Let $L$ be a regular language over the finite language 
$\Gamma$ with minimal automaton $M(L)$ having states $Q$. Let 
$Q^{acc}$ be the set of accepting states of $M(L)$, and let 
$q_{ini}$ be the initial state of $M(L)$. Because $M(L)$ is the 
minimal automaton for $L$, 
 all states in $Q$ are reachable from $q_{ini}$; that is, there exists $a\in \Gamma^*$ such that $q_{ini}.a= q$.

If $L$ is presented by a DFA $F$, then $M(L)$ can be determined in 
polynomial time (with instance size the number of states of $F$). If a regular language $L$ is presented as a regular expression 
$\mathcal{E}$, or by an NFA $N$, then $M(L)$ can be determined in 
EXPSPACE from $\mathcal{E}$ and $N$, respectively.

 For $q\in Q$, let $L_q$ 
be the language determined by the automaton having the same 
transition function and final states of $M(L)$, but with $q$ (rather 
than $q_{ini}$) as the initial state. Let $|M(L)|$ be the number of 
states of $M(L)$. In the proof of Lemma~\ref{limits cellular}, use 
will be made of the fact that the state set $Q$ of $M(L)$ can be 
taken to be $\{w/\sim_L: w \in \Gamma^*\}$, where $\sim_L = 
\{(u,v)\in \Gamma^*: \forall w \in \Gamma^*, \; uw\in L 
\leftrightarrow vw \in L\}$. Note that $\sim_L$ is a right 
congruence of finite index of the free monoid $\Gamma^*$, where the 
action of $\Gamma$ on $Q$ is given by $w/\sim_L \to w.c/\sim_L$, 
where $c\in \Gamma$. The final states are those $\sim_L$ classes 
contained in $L$; the initial state is $\epsilon/ \sim_L$, where 
$\epsilon$ is the empty word.

\vspace{.1 in}

Given an NFA $N$, let $|N|$ be the number of its states. For NFAs 
$A$ and $B$, it is well known that determining whether $L(A) 
\subseteq L(B)$ (and thus whether $L(A) = L(B)$) is in PSPACE with 
respect to an instance size of $O(|A|+|B|)$.

\section{Proof of Theorem~\ref{main result}}

In the folklore result below, regular languages that satisfy (L1) 
are characterized in terms of two conditions on their minimal 
automata; a deterministic automaton satisfying the first condition 
below is sometimes referred to as a {\bf sink automaton}.

\begin{lemma} \label{limits cellular} If $L\neq \Gamma^*$, then 
$L$ satisfies (L1) if and only if the minimal automaton $M(L)$ of 
$L$ satisfies:
 \begin{enumerate}
\item Among the states $Q$ of $M(L)$, there is a unique non-accepting state 
$g$; moreover $g$ is a sink.

\item If $q\in Q- \{g\}$, the language
 $L_q$ is contained in $L$.
\end{enumerate}
\end{lemma}
{\bf Proof.} Suppose $L \neq \Gamma^*$ and $L$ satisfies (L1). Let 
$u,v \in \Gamma^*$. If $u \not \in L$ and $v \not \in L$, then for 
all $c\in \Gamma^*$, by (L1), we have both that $uc \not \in L$ and 
$vc \not \in L$. By the definition of $\sim_L$, we have $u$ and $v$ 
are in the same class of $\sim_L$. It follows that there is only one 
non-accepting state, the class $u /\sim_L$, where $u$ is any word 
not in $L$. Denote the non-accepting state by $g$. By (L1) again, it 
follows that $g$ is fixed by each element of $\Gamma$; that is, $g$ 
is a sink.

\vspace{.1 in}

Suppose $q\in Q- \{g\}$. It is shown that $L_q \subseteq L$. Since 
$M(L)$ is the minimal automaton for $L$, $q$ is reachable from 
$q_{ini}$. So suppose $d\in \Gamma^{\ast}$ is such that $q_{ini} .d 
= q$. Let $c\in L_q$. Then $q.c \neq g$, and $q_{ini}.dc \neq g$, 
from which it follows that $dc\in L$. Now (L1) guarantees that $c\in 
L$. \vspace{.1 in}

Conversely, assume $L \neq \Gamma^*$ and the two conditions above 
hold for $L$ and $M(L)$. Suppose $uv\in L$, where $u,v \in 
\Gamma^*$. That $M(L)$ is a sink automaton implies that $q_{ini}.u 
\in Q^{acc}$; that is, $u\in L$. Now using that $L_u \subseteq L$, 
it follows that $v\in L$. $\square$

\vspace{.1 in}

 Given $M(L)$, determining if $L$ 
satisfies (L1) amounts to checking that the non-accepting state 
is a sink, and for each $q\in Q^{acc}$ checking that $L_q$ is 
contained in $L$. This last test is in PSPACE with respect to 
instance size $O(|M(L)|)$. Also, given $M(L)$, it is obviously in 
$P$ to check that the single non-accepting state is a sink. The 
following 
 has been proved. For a regular expression $\mathcal{E}$, let 
$|\mathcal{E}|$ be the number of symbols in $\mathcal{E}$. 

\begin{lemma}\label{(L1) NL} Given a regular language $L$ presented by a regular expression $\mathcal{E}$ or as a 
non-deterministic automaton $N$, then the problem with 
question``Does $L$ satisfies (L1)?'' is decidable; indeed, it is in 
EXPSPACE with respect to $|N|$ or $|\mathcal{E}|$. 
\end{lemma}

{\bf Proof of Theorem~\ref{main result}.} 
To prove Theorem~\ref{main result}, it suffices to show that given a regular language $L$ satisfying (L1) and its minimal automaton 
$M(L)$, whether (L2) is satisfied can be determined algorithmically. 
Of course to show that (L2) is decidable, it suffices to show that 
given $o\in \Gamma$, it can be decided whether (L2) with $o$ is 
satisfied. For the remainder of the proof, $o$ will be a fixed 
element of $\Gamma$. From Lemma~\ref{limits cellular}, it can be 
assumed $M(L)$ is a sink automaton, and all states of $M(L)$ are 
reachable from $q_{ini}$.

\begin{definition} \label{o-paths}
Let $q\in Q^{acc}$ be such that for all $n\in \mathbb{N}$, $q.o^n$ 
is in $Q^{acc}$, and let $P$= $q.o^*= \{q.o^n: n\in \mathbb{N}\cup 
\{0\}\}$; $P$ 
 will be treated as a simple path that 
begins with $q$, and $P$ will be said to be an {\bf o-path}.

Of course, there are no more than $|M(L)|$ $o$-paths. 

\end{definition}

\begin{definition} Let $q\in Q^{acc}$, and let $P(o,q)$ be the set of all $o$-paths $P$ for 
which there exists a directed path from $q$ to the first state of 
$P$.
\end{definition}

After some motivation, for $q\in Q^{acc}$, a regular language 
$L(o,q)$ will be defined: 
 Suppose that $L$ satisfies (L2) with $o$. Let $x,y$ be contained in $L$. 
 Since $L$ satisfies (L1) and (L2), independent of $n$ there exist $a,b\in L$ such that 
$xao^nby\in L$. Let $q = q_{ini}.x\in Q^{acc}$. Since $L$ satisfies 
(L1), for all $n\in \mathbb{N}$, we have $q.ao^n \in Q^{acc}$. Thus 
$P= q.ao^*$ is an $o$-path. Indeed, $P$ is in $P(o,q)$. From 
$xao^nby\in L$, it follows that for all $p\in P$, the regular 
language $L_{p.b}$ contains $y$. That is, $y\in \cap_{p\in P} 
L_{p.b}$. Thus, if $L$ satisfies (L2) with $o$, then for all $q\in 
Q^{acc}$ and all $y\in L$, there exists an $o$-path $P$ in $P(o,q)$ 
 and a word $b\in L$ such that $y\in \cap_{p\in P} \;
L_{p.b}$. 

Ignoring $q$ for the time being, notice for a given $y\in L$ and 
$o$-path $P$, to check whether 
 such a $b\in 
L$ exists, it suffices to check only those $b\in L$ having length no 
more than $|M(L)|^{|M(L)|}$; this is because $b$ only matters here 
up to its action on $M(L)$. Related to this observation, and 
following for the same reason, if $L$ satisfies (L2), given $x,y\in 
L$, independent of $n$, there exist $a,b\in L$ such that $|a|,|b| 
\leq |M(L)|^{|M(L)|}$ and $xao^nby\in L$.

\begin{definition} \label{L(o,q)}
We introduce the following notation for the sake of brevity.
\begin{enumerate}

\item Let $K(o,q) = P(o,q) \times (\Gamma^{(|M(L)|^{|M(L)|})})$.

\item Let $L(o,q) = \cup_{(P,b)\in K(o,q)} (\cap_{p\in P} L_{p.b})$.
\end{enumerate}

\end{definition}

Note that since $L$ satisfies (L1), it follows that $L(o,q)$ is contained 
in $L$, and since $L(o,q)$ is a finite Boolean combination of 
regular languages, $L(o,q)$ is regular. For any $q\in Q^{acc}$, if 
$L(o,q)$ is empty, observe that (L2) can not be satisfied with $o$. So it is harmless to assume $L(o,q)$ is non-empty.

\begin{proposition}\label{main proposition} $L$ satisfies (L2) with $o$ if and only if for all $q\in Q^{acc}$, we have 
$L(o,q) = 
L$. 
\end{proposition}
{\bf Proof.} Suppose that for all $q\in Q^{acc}$ we have $L = 
L(o,q)$. Let $x,y$ be in $L$ and let $q = q_{ini}.x$. Since $y\in L= 
L(o,q)$, there exists $P\in P(o,q)$, $b\in L$ with $|b|\leq 
|M(L)|^|M(L)|$ such that $y\in \cap_{p\in P} L_{p.b}$. Let $p_1$ be 
the first state of $P$. Since $P\in P(o,q)$, there exists $a\in L$ 
such that $q.a= p_1$. But now we have for $n\in \mathbb{N}$ that 
$q_{ini}.xao^nby\in Q^{acc}$; equivalently, we have $xao^nby\in L$. 
\emph{Note that $a,b$ are both independent of $n$.}

Conversely, suppose that $L$ satisfies (L2) with $o$. Let $q\in 
Q^{acc}$. We need to show that $L = L(o,q)$. Since $M(L)$ is the 
minimal automaton of $L$, there exists $x\in L$ such that $q= 
q_{ini}.x$. Select arbitrary $y\in L$. We will show that $y\in 
L(o,q)$. Since $L$ satisfies (L2) with $o$, there exist $a,b\in L$ 
such that for any $n\in \mathbb{N}$ we have $xao^nby\in L$. 
Moreover, we can assume that $|b| \leq |M(L)|^{|M(L)|}$. Since $L$ 
satisfies (L1), we have for all $n\in \mathbb{N}$ that $xao^n\in L$; 
thus, $q_{ini}.xao^n \in Q^{acc}$, and $q.ao^*$ is an $o$-path $P$ 
having first state $q.a$. Moreover, since $P$ is reachable from $q = 
q_{ini}.x$, we have $P\in P(o,q)$. But now $xao^nby\in L$ implies 
that $y\in \cap_{p\in P} L_{p.b}$, where $P \in P(o,q)$ and $|b| 
\leq |M(L)|^{|M(L)|}$. That is, $y\in L(o,q)$. $\square$

\vspace{.1 in}
 
{\bf Completion of proof of Theorem~\ref{main result}.} To prove the 
theorem, it suffices to show for a given $q\in Q^{acc}$ that 
determining whether $L = L(o,q)$ is decidable. Using the determinism 
of $M(L)$, the collection $P(o,q)$ can be constructed 
algorithmically. Decidability is proven if the Boolean combination 
 of languages 
of the form $L_{p.b}$ can be constructed algorithmically. We can 
algorithmically form a non-deterministic automaton $N$ such that 
$L(N)= L(o,q)$, where $|N|$ is bounded above by a recursive function 
in $|M(L)|$: Letting $\alpha = |M(L)|$, 
 by examining the parts of 
the Boolean expression determining $L(o,q)$, and using the standard 
constructions for non-deterministic machines that represent unions 
and intersections of regular languages, we can bound $|N|$ as 
follows. Let $\alpha = |M(L)|$. Observe that $|K(o,q)| \leq \alpha 
(\alpha^{\alpha})$, and that for $(P,b)\in K(o,q)$, we can determine 
$\cap_{p\in P} L_{p.b}$ with a non-deterministic automaton having no 
more than $\alpha^2$ states. Thus to construct an NFA that accepts 
$L(o,q)$, we need no more than $2(\alpha) \alpha^{\alpha} \alpha^2$ 
states. It follows now that determining whether $L$ is equal to 
$L(o,q)$ is decidable, completing the proof of Theorem~\ref{main 
result}. $\square$

\end{document}